\documentclass[a4paper]{spie}  
\usepackage[]{graphicx}

\title{
Development of frequency domain multiplexing for the X-ray Integral Field Unit (X-IFU) on the Athena.
} 

\author{
Hiroki Akamatsu, Luciano Gottardi, Jan van der Kuur, Cor P. de Vries, Kevin Ravensberg, \\
Joseph S. Adams, Simon R. Bandler, Marcel P. Bruijn, James A. Chervenak, \\
Caroline A Kilbourne, Mikko Kiviranta, A.J. van den Linden, Brian D. Jackson and,  \\
Stephen J. Smith 
\skiplinehalf
\supit{a}SRON Netherlands Institute for Space Research, Sorbonnelaan 2, 3584CA Utrecht, The Netherlands\\
\supit{b}SRON Netherlands Institute for Space Research, Landleven 12, 9747AD, Groningen, The Netherlands\\
\supit{c}NASA	Goddard Space Flight Center.  662	Greenbelt, MD, USA \\
\supit{d}VTT, Tietotie 3, 02150 Espoo, Finland
}

\authorinfo{
Corresponding author: H. Akamatsu \\
E-mail: h.akamatsu@sron.nl, Telephone: +31 88 777 5862}

\begin{document} 
  \maketitle 

\begin{abstract}
We are developing the frequency domain multiplexing (FDM) read-out of transition-edge sensor (TES) microcalorimeters for the X-ray Integral Field Unit (X-IFU) instrument on board of the future European X-Ray observatory Athena. The X-IFU instrument consists of an array of $\sim$3840 TESs with a high quantum efficiency ($>$90 \%) and spectral resolution $\Delta E$=2.5 eV $@$ 7 keV ($E/\Delta E\sim$2800). 
FDM is currently the baseline readout system for the X-IFU instrument. 
Using high quality factor LC filters and room temperature electronics developed at SRON and low-noise two stage SQUID amplifiers provided by VTT, we have recently demonstrated good performance with the FDM readout of Mo/Au TES calorimeters with Au/Bi absorbers.  An integrated noise equivalent power resolution of about 2.0 eV at 1.7 MHz has been demonstrated with a pixel from a new TES array from NASA/Goddard (GSFC-A2).
We have achieved X-ray energy resolutions $\sim$2.5 eV at AC bias frequency at 1.7 MHz in the single pixel read-out. We have also demonstrated for the first time an X-ray energy resolution around 3.0 eV in a 6 pixel FDM read-out with TES array (GSFC-A1). In this paper we report on the single pixel performance of these microcalorimeters under MHz AC bias, and further results of the performance of these pixels under FDM. 
\end{abstract}



\section{INTRODUCTION}
\label{sec:intro}  
Future X-ray astronomical satellite {\it Athena}\cite{nandra13} (2028$\sim$) aims to unveil the hot and energetic side of the Universe. In order to accomplish the goal, {\it Athena} will employ two focal plane instruments such as Wide Field Imager (WFI\cite{wfi14}) and X-ray Integrated Field Unit (X-IFU\cite{barret13, xifu14}). The X-IFU instrument will provide a superb X-ray spectral ($\sim 2.5 eV~< 7$ keV) and spatial ($\sim 5^{''}$) resolutions.
Transition edge sensors (TESs) X-ray calorimeter is a current baseline of the X-IFU instrument. X-ray calorimeter is a cryogenic non-dispersive spectrometer. TESs use a sharp resistance drop of a superconducting film as a thermometer operated around 100 mK.  With high sensitive thermistor TESs, X-ray calorimeter can archive superb spectral resolution\cite{mmm84} ($\sim$ 1 eV $@$ 6 keV). Furthermore, because of a non-dispersive spectrometer, TESs X-ray calorimeter can be used for diffuse objects such as super nova remnants, galaxies and galaxy clusters.

Although TESs X-ray calorimeter will innovate X-ray spectroscopy of cosmic plasma, the instrument needs to satisfy severe constraints on the satellite (the electrical and the cooling power). Therefore, a multiplexing readout of the TESs X-ray calorimeter is crucial technology. SRON is developing the frequency domain multiplexing (FDM) readout. 
In the FDM, TESs are coupled to a passive LC filter and biased with alternating current (AC bias) at MHz frequencies. Each LC resonator should be separated beyond detector thermal response ($<$ 50 kHz) to avoid crosstalk between neighboring resonators.  
To satisfy XIFU requirements, a multiplexing factor of 40 pixels/channel in a frequency range from 1 to 5MHz required. The detailed description of the bandwidth requirement of FDM  is given in J. van der Kuur et al. 2016~\cite{jan16_SPIE}. 
In this paper, we report on our recent progress in the development of the SRON FDM read-out for a  NASA/GSFC TESs calorimeter array.

\section{NASA/GSFC TES microcalorimeter array}\label{sec:TESs}
For the FDM demonstration, we are using two different NASA/GSFC TESs arrays:
(1) 8$\times$8 uniform array (GSFC-A1) and (2) Mixed array (GSFC-A2). The basic properties of these array are summarized in Table.~\ref{tab:gsfc}. 
Both the arrays show an excellent performance under the DC bias, typically 1.8--2.4 eV. 
The TESs consist of thin Mo/Au bilayer films and have 250 $\times$ 250 um BiAu-mushroom absorbers. Both TESs array have an Au layer on top of Si substrate to reduce thermal crosstalk and the bath temperature fluctuation due to X-ray photon attack onto the Si.
The basic properties of GSFC-A1 are reported by C. Kilbourne 2007~\cite{kilbourne07}, Iyomoto et al. 2008~\cite{iyomoto08}, etc.
GSFC-A2 has different absorber connections and wiring configuration. The absorbers are connected to TESs via T-type stem structures. The wiring of GSFC-A2 is made of the strip-line to increase filling factor and reduce electrical cross-talk. GSFC-A2 also has an better thermalization layer than GSFC-A1, which reduces thermal crosstalk in the array.

\begin{figure}[h]
\begin{tabular}{c}
\begin{minipage}{1.\hsize}
\begin{center}
\includegraphics[width=.45\hsize]{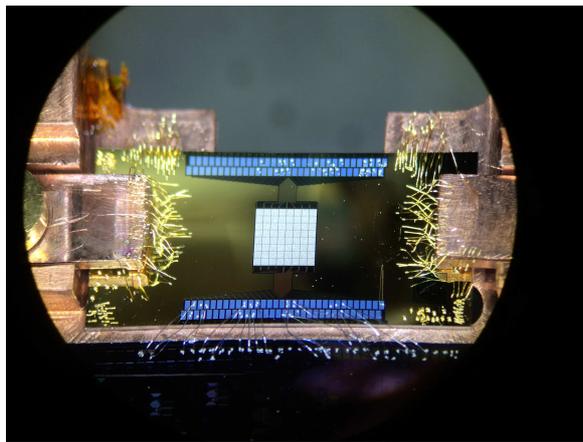}
\end{center}
\end{minipage}
\end{tabular}
\caption{\label{fig:gsfc-a2}
Zoomup picture of $8\times8$ TESs GAFC-A2. TESs chip is clamped by Cu bars and thermally connected with Au wire bondings.
}
\end{figure}

The TES arrays are  clamped by Cu bars and thermally coupled to the Cu bracket via several Au bonding (Fig.\ref{fig:gsfc-a2}).
With GSFC-A1, we have improved our FDM readout system and investigated the detector performance under AC bias (Akamatsu et al. 2013\cite{akamatsu13_ASC12}, 14\cite{akamatsu14_ACBias}, 15\cite{akamatsu15_ASC14} and Gottardi et al. 2012\cite{gottardi12_LTD}, 14\cite{gottardi14_SPIE}, 16\cite{gottardi16}).

\begin{table}[htdp]
\caption{Basic properties of GSFC TESs calorimeter array\label{tab:gsfc}}
\begin{center}
\begin{tabular}{ccccccccccccccc} \hline
		&	TES	size& 	Absorber 	& 		Transition 		 & Normal state	&	Saturation 	&Temperature \\
		&	[$\mu$m$^2$]	&size [$\mu$m$^2$] &	 temperature	 [mK] & resistance [m$\Omega$]	& power [pW] &sensitivity $\alpha^\ast$	\\ \hline
GSFC-A1	& 	140			&	250 			&	95	& 	7.5	&	$\sim$6.5		&	60	\\ 
GSFC-A2	&	100			&	250			&  	93	&	8.3	&	$\sim$4.7		&	70	\\
		&	120			&	250			&	95 	&	9.0	&	$\sim$6.1		&	70	\\
		&	140			&	250			&	97	&	9.8	&	$\sim$7.2		&	80 	\\
		\hline
\multicolumn{7}{l}{$\ast$: Dimensionless temperature sensitivity of the thermistor $\alpha\equiv\frac{{\rm d ln} R}{{\rm d ln} T}$}\\
\end{tabular}
\end{center}
\end{table}%

\begin{figure}[h]
\begin{tabular}{c}
\begin{minipage}{.6\hsize}
\begin{center}
\includegraphics[width=1.\hsize]{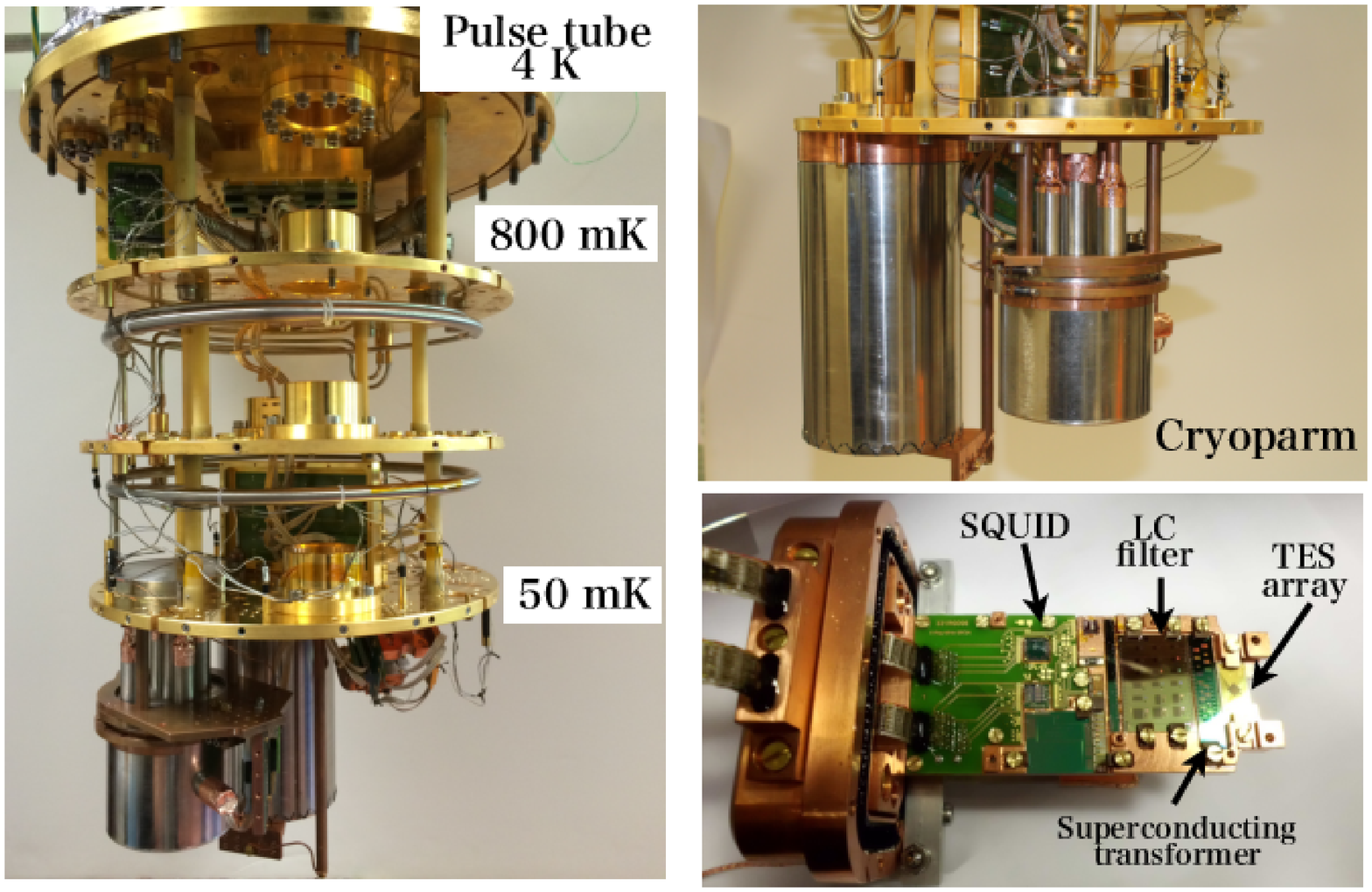}
\end{center}
\end{minipage}
\begin{minipage}{.37\hsize}
\begin{center}
\includegraphics[width=.8\hsize]{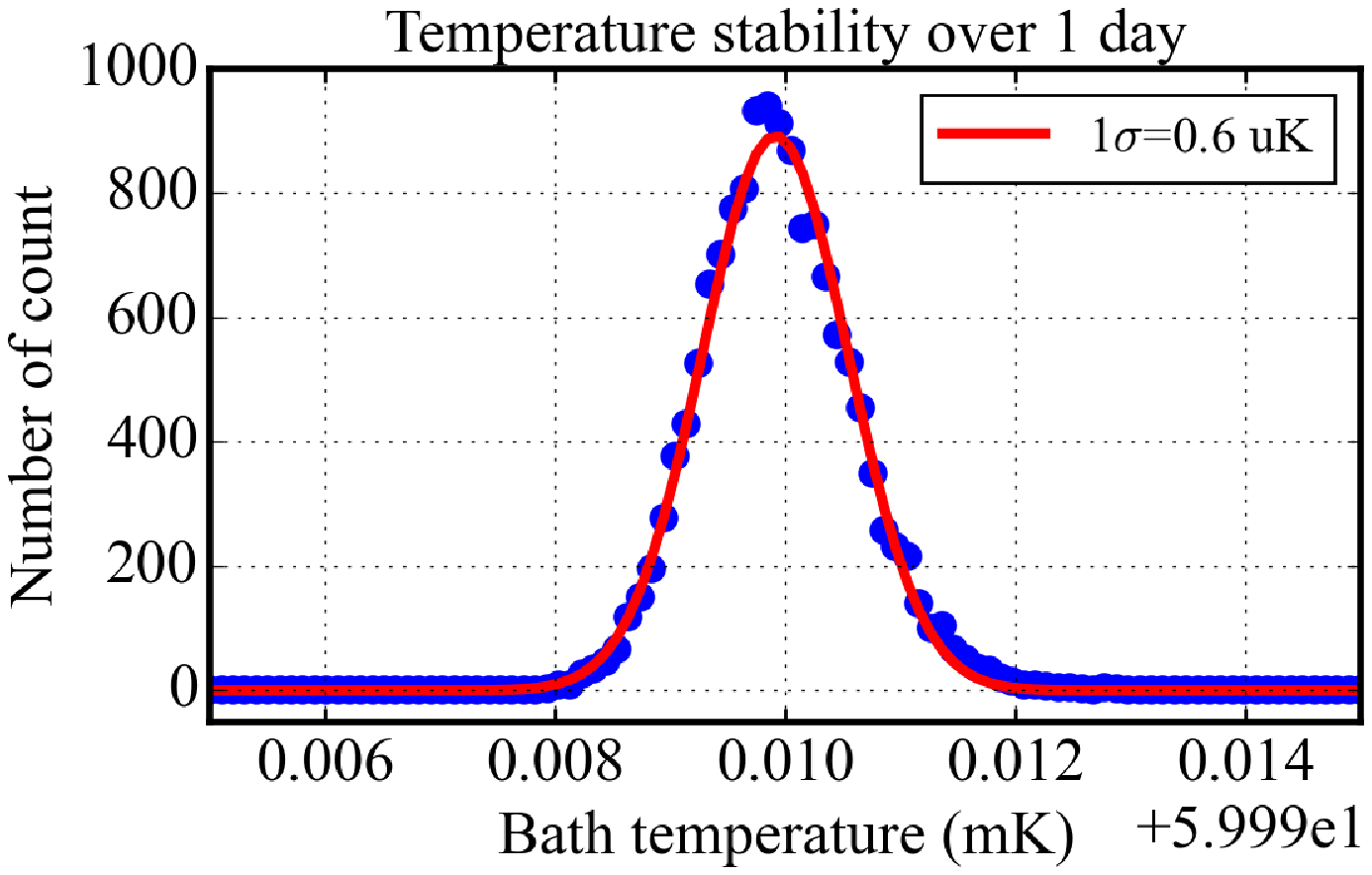}
\includegraphics[width=.8\hsize]{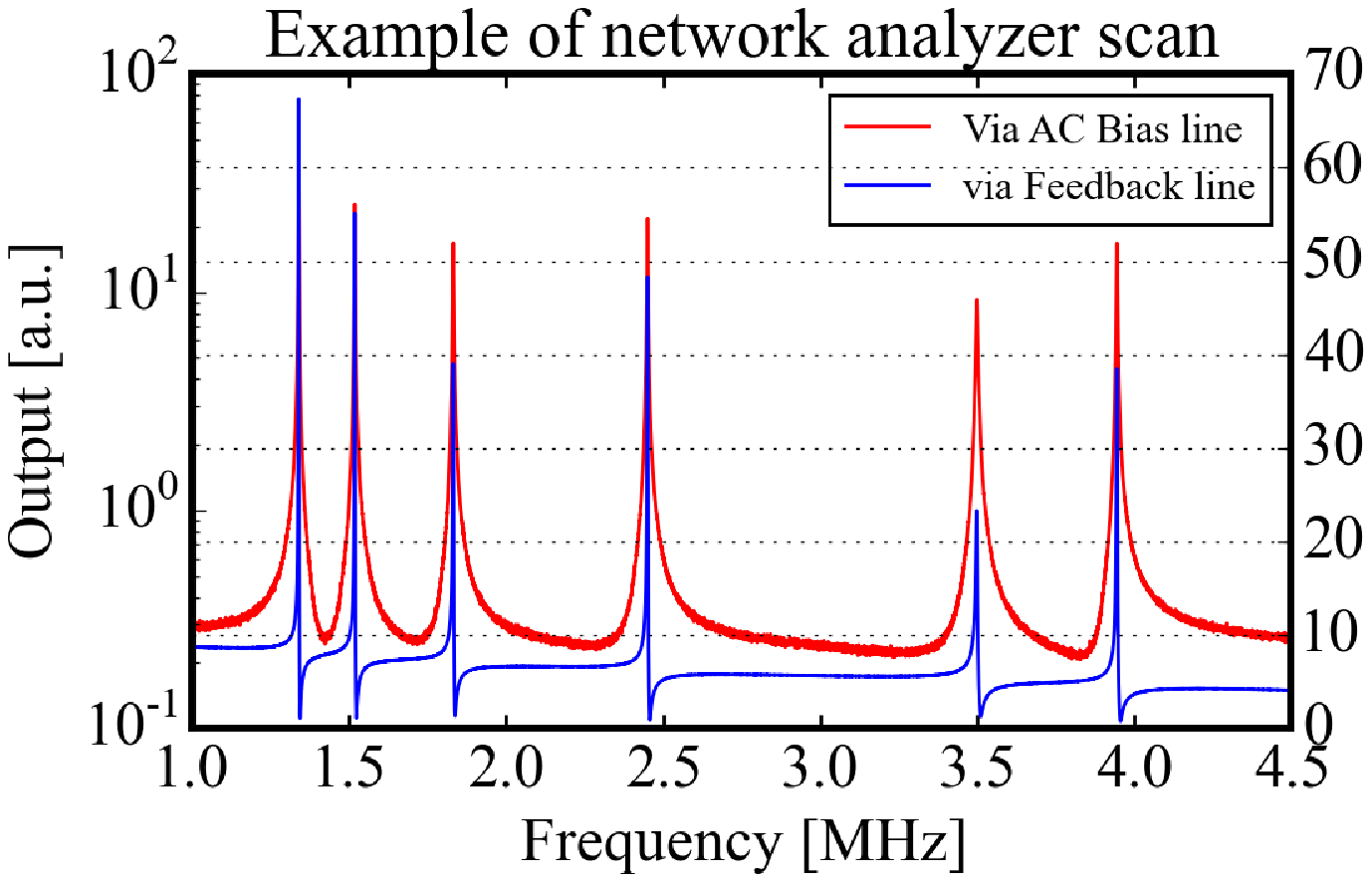}
\end{center}
\end{minipage}
\end{tabular}
\caption{\label{fig:setup}
Left and middle: Pictures of the FDM set-up.
Right top: Example of the bath temperature stability. Throughout a day, the bath temperature is reasonably stable with $\sigma=0.6~\mu \rm K$ at 60 mK.
Right bottom: Example of the network analyzer scan. We connected 6 GSFC TESs calorimeters  in series with the LC filters with resonance frequencies of 1.3, 1.5, 1.7, 2.4, 3.4 and 3.9 MHz respectively. Red and Blue indicate the network analyzer scan via AC bias and Feedback line, respectively.
}
\end{figure}

\section{EXPERIMENTAL SETUP}\label{sec:setup}
{\bf Cryostat:} For the FDM demonstration, we are using a cryogen free dilution cooler\footnote{http://www.leidencryogenics.com/} (Fig.~\ref{fig:setup}).  The cooler has a huge cooling power $\sim$ 400 $\mu$W$@$ 100 mK, which hosts several setups. Currently 2 FDM set-ups with s X-ray calorimeter array are installed. We have developed stable magnetic field and light tight set-up\cite{gottardi14_SPIE}. We employed a high-$\mu$ metal Cryoperm shield and superconducting Nb shield. The Nb shield cools down from one point from the detectors side to control flux trapping. Furthermore, we introduced a Helmholtz coil to investigate TESs response as a function of the applied magnetic field. 
With a similar setup, we have demonstrated an ultra-low NEP (Noise equivalent power) bolometer for the first time ($NEP_{\rm dark}\sim 1\times10^{-19}\rm ~W/\sqrt{Hz}$: Suzuki et al. 2016\cite{suzuki16}). 
For the performance evaluation of TESs calorimerter, stable bath temperature is of importance. We employed a highly sensitive Ge thermistor, which has a temperature sensitivity $\alpha (\equiv\frac{{\rm d ln} R}{{\rm d ln} T})\sim5$ at 50 mK.

\noindent
{\bf Room temperature electronics:} SRON is developing room temperature electronics for the FDM readout.  The details are summarized in den Hartog et al. 2009\cite{hartog09}.

\noindent
{\bf SQUIDs:}
We are using low-noise two-stage SQUID amplifiers  provided by VTT\footnote{http://www.vttresearch.com/}.
The SQUID amplifiers are mounted on the Cu bracket and cooled down together with other experimental components. 
For the single pixel characterization,  we employed SQUIDs, which are nearly quantum-limited with an coupled energy resolution $\sim20\hbar$ at 20 mK. The SQUID input current noise  shows 1-2 pA/$\sqrt{\rm Hz}$ over the required frequency range  between 1--5 MHz as expected for these SQUID amplifiers. The detailed information about the VTT SQUID amplifiers can be found in L. Gottardi et al. 2015\cite{gottardi15_SQUID}.
 For the multiplexing demonstration, we employed a higher dynamic range SQUIDs  at the cost of  slightly higher  noise 4-6 pA/$\sqrt{\rm Hz}$.
 
 \noindent
{\bf Superconducting transformer:}
We employed a superconducting transformer to 
match the read-out  impedance to the low ohmic impedance ($\sim 7-10~\rm m\Omega$) of the GSFC TESs calorimeter (Tab~\ref{tab:gsfc}) and optimize the SQUID dynamic range. In our setup, the superconducting transformer works as to match the SQUID dynamic range and the impedences between TESs and SQUID.
In this paper, we are using SRON lithographic  superconducting transformers with a coupling ratio $n=5$ or 8.

\noindent
{\bf LC filter:} SRON developed  low-losses lithographic LC filters for the FDM readout. The LC filter made of a-Si:H with gradiometric geometry and strip-line wiring to reduce common impedance and mutual inductance. The nominal inductance of the coil used in each filter is $L$=400 nH (GSFC-A2) or 2 $\mu$H (GSFC-A1).
The detailed information about the LC filter can be found M. Bruijn et al. (2012)\cite{bruijn12}.

\begin{figure}[t]
\begin{tabular}{c}
\begin{minipage}{.52\hsize}
\begin{center}
\includegraphics[width=1.\hsize]{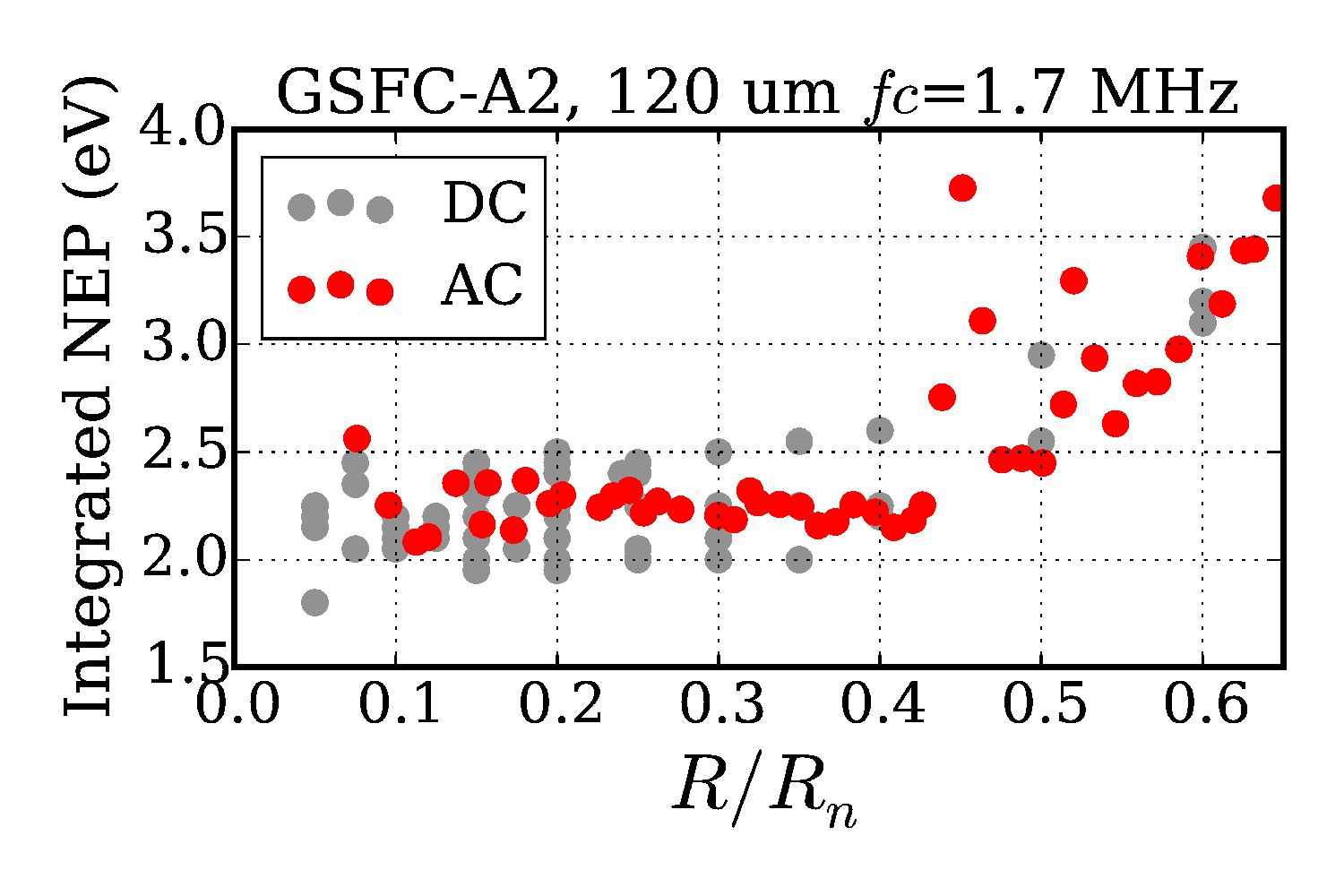}
\includegraphics[width=1.\hsize]{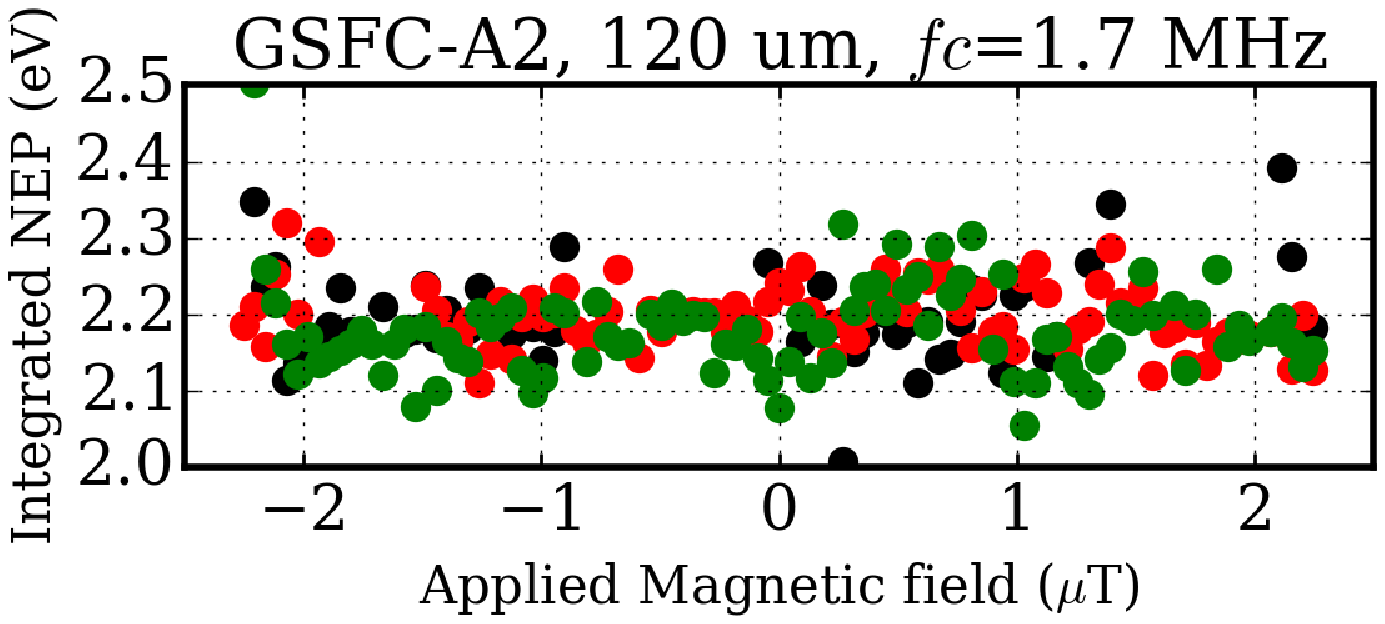}
\end{center}
\end{minipage}
\begin{minipage}{.48\hsize}
\begin{center}
\includegraphics[width=1.\hsize]{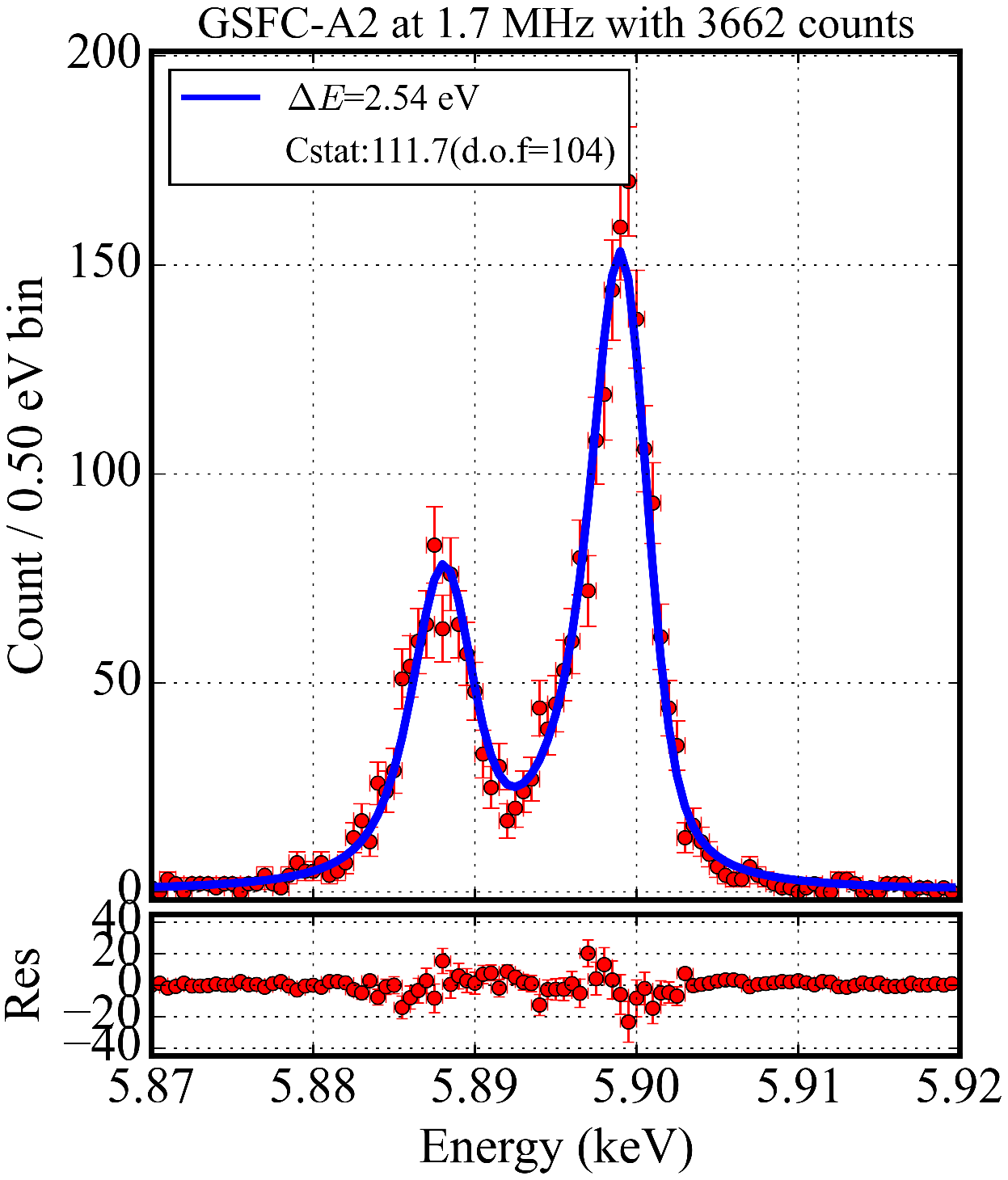}
\end{center}
\end{minipage}
\end{tabular}
\caption{\label{fig:AC_single}
Left top: the integrated NEP resolution as a function of the normalized TES resistance. Red and gray points represent 120 $\mu$m pixel of GSFC-A2 with 1.7 MHz AC bias and DC bias, respectively.
Left bottom:  the integrated NEP resolution as a function of applied magnetic field.
Right: Energy spectrum of Mn-K$\alpha$ X-rays. The data is shown in red crosses. The blue curves show the best fit model.
}
\end{figure}

\section{RESULTS}\label{sec:results}
Here we present the results of  the single pixel characterization of GSFC-A2 at 1.7 MHz AC bias.
The 120 $\mu$m TES calorimeter pixel was connected to a 1.7 MHz LC resonator ($L=$400 nH) with a 1:8 superconducting transformer. We used the $^{55}$Fe X-ray source, which was mounted on the Nb magnetic shield and cooled down together. To avoid X-rays hitting the Si substrate, we employed a Cu collimator with a hole, which fits the size of the TES array. Typical count rate to TESs is about 1.0--1.5 counts/s.

In order to evaluate the detector performance under AC bias, we characterized the integrated NEP resolution.
The integrated NEP resolution reflects a potential performance under given measurement set-up. 
The integrated NEP resolution was estimated based on below formula
\begin{equation}
\Delta E_{NEP}=\left(\int \frac{4df}{ NEP(f)}\right)^{-1},
\end{equation}
where $NEP(f)\equiv\frac{e_n(f)}{S_v(f)}$, $e_n(f)$ and $S_v(f)$ are the detector noise spectral density and responsivity, respectively.
The top panel of Fig.~\ref{fig:AC_single} shows the dependency of the NEP resolution on the TES bias point (red points). The horizontal axis is normalized on the normal state resistance. 
The gray points show the results of DC bias. Within a scatter of the data, the integrated NEP resolutions under AC bias are almost comparable to the DC bias case. The profile shows a stable NEP resolution as $2.0-2.3$ eV between TES resistance $R=0.1-0.4~R_{\rm N}$ and the best  resolution as $\sim$ 2.0 eV at $R=0.12~R_{\rm N}$.
As previously reported~\cite{akamatsu15_ASC14},  the degradation of the integrated NEP resolution under AC bias has been observed at small TES resistance regime ($R<0.25~R_{\rm N}$) with GSFC-A1 array. On the other hand, there is no significant difference between AC and DC\cite{smith_16SPIE} bias with the GSFC-A2 array. The TES parameters (Tab.~\ref{tab:gsfc}), the magnetic field sensitivity (Fig.~\ref{fig:AC_single}) and the impact of the weak-link effect~\cite{sadleir11_WL, smith13_wl, gottardi14_WL} may be  responsible for the difference. 


To assess the X-ray resolution of the TESs calorimeter,  we applied the optimal filter to the X-ray pulses.
After the drift and non-linearity correction, we fitted Mn-K$_\alpha$ line with a line model by Holzer et al (1997)~\cite{holzer97}. For the fitting, we employed the Cash statistic\cite{cash79} to minimize fitting bias (see the SPEX user manual\footnote{https://www.sron.nl/files/HEA/SPEX/manuals/manual.pdf} chapter 2.12 for a detail). 
The best-fit parameter then was obtained by minimizing the C-stat parameter. The best X-ray energy resolution under 1.7 MHz AC bias is $\Delta E=2.54$ eV at 5.9 keV, which is close to  typical values under DC bias (2.0--2.4 eV). The difference between the integrated NEP and X-ray resolution could be cause by thermal and mechanical fluctuations induced by the external environment. This effect is currently under investigation.

Finally we briefly report on a preliminary result of 6 pixel FDM demonstration. For the FDM demonstration, we used relatively stable (temperature and magnetic field) set-up, which hosts GSFC-A1 array.  
As described in Sec~\ref{sec:TESs}, the GSFC-A1 array is connected 2$\mu$H coil LC filter.
We connected 8 TES calorimeters, to LC filters with resonance frequencies of 1.10, 1.27, 1.38, 1.55, 1.75, 2.05, 2.45 and 2.55 MHz respectively.  Because of undesired detector behaviors (too fast detector response), the TESs connected to 1.10 and 1.38 MHz are excluded from the multiplexing measurement.  With this condition, we demonstrated 6 pixel multiplexed read out with typical energy resolutions of $\sim$ 3 eV for the first time.  

Contrary to the previous 2-pixel multiplexing\cite{akamatsu16_ltd15}, there is a small performance degradation as a result of 6-pixel multiplexing from $\sim2.8$ eV (single pixel mode) to $\sim$ 3.2 eV (6-pixel multiplexing). The degradation can be explained by sub-optimal components such as TES array and LC filter. 
The bias line layout of the TES array used for multiplexing experiment showed excess cross talk.
Consequently, undesired electrical cross-talk is generated. Furthermore, the thermalization efficiency of the array is an issue. We observed strong thermal cross-talk between TESs and a sign of X-ray hit on the Si substrate. We also observed an electrical cross-talk which is most likely related to a common impedance and mutual inductance  in this specific LC filter version. The degradation will be improved by the new generation of LC filters with minimised common inductance and mutual inductance and by the new generation of detector currently under fabrication at NASA/Goddard. In parallel, the demonstration of the FDM readout is still ongoing and the results will be reported in the near future.


\section{Summary and Future prospect}\label{sec:future}
We are developing the Frequency Domain Multiplexing readout of TESs calorimeter for the X-IFU onboard the future X-ray astrophysical satellite $Athena$.
By employing a new TES calorimeter array, we have demonstrated  an integrated NEP resolution of $2.0-2.3$ eV under AC bias at 1.7 MHz, which is consistent with the results of DC bias measurements. The energy resolution of $\sim 2.5$ eV at 5.9 keV with the single pixel MHz AC bias readout is also presented.

\begin{figure}
\begin{tabular}{c}
\begin{minipage}{1\hsize}
\begin{center}
\includegraphics[angle=-90, width=.5\hsize]{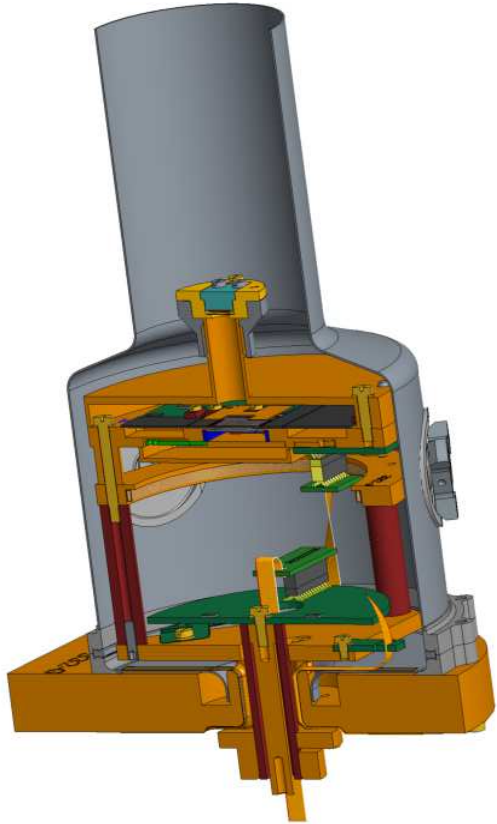}
\end{center}
\end{minipage}
\end{tabular}
\caption{CAD image of 40$\times$2 pixel demonstrator. \label{fig:40pix}}
\end{figure}

For the near future we are preparing a new experimental setup, which is shown in Fig.~\ref{fig:40pix}.
This setup is design to test $2\times40$ pixels FDM readout. For this demonstration, we will employ (1) a uniform GSFC TESs array with will have similar properties of GSFC-A2, (2) new LC filter and (3) X-IFU dedicated SQUID array.  The first cool down of the set up is expected to be around early winter of 2016.

\appendix    

\acknowledgments     
H.A acknowledges the support of NWO via a Veni grant. 
We thank Martijn Schoemans, Dick Boersma, Marcel van Litsenburg, Joris van Rantwijk, Patrick van Winden and Bert-Joost van Leeuwen and Robert Huitin for their precious/continues support.
The research leading to these results has received funding from the
European Union’s Horizon 2020 Programme under the AHEAD project (grant
agreement n. 654215).
SRON is supported financially by NWO, the Netherlands Organization for Scientific Research.

\bibliography{report}   
\bibliographystyle{spiebib}   

\end{document}